\documentclass[conference]{IEEEtran}
\IEEEoverridecommandlockouts
\usepackage{cite}
\usepackage{amsmath,amssymb,amsfonts}
\usepackage{algorithmic}
\usepackage{listings}
\usepackage{breqn}
\usepackage{enumitem}

\usepackage{graphicx}
\usepackage{subcaption}
\usepackage{textcomp}
\usepackage{xcolor}
\usepackage{url}
\def\BibTeX{{\rm B\kern-.05em{\sc i\kern-.025em b}\kern-.08em
    T\kern-.1667em\lower.z7ex\hbox{E}\kern-.125emX}}
{\renewcommand{\baselinestretch}{0.95}

\begin{document}
\lstset{
    language=Python,
    basicstyle=\ttfamily\footnotesize,
    breaklines=true,
    postbreak=\mbox{\textcolor{red}{$\hookrightarrow$}\space},
    showstringspaces=false,
    numbers=none,
    commentstyle=\color{green!50!black},
    keywordstyle=\color{blue},
    stringstyle=\color{red},
    breakatwhitespace=false,
    keepspaces=true,
    columns=flexible,
    framexleftmargin=1em,
    xleftmargin=1em,
    backgroundcolor=\color{gray!10},
    tabsize=4,
}
\title{LoRaFlow: High-Quality Signal Reconstruction using Rectified Flow%
}

\author{Mohamed Osman \\
\IEEEauthorblockA{\textit{Department of Computer Science} \\
\textit{Virginia Commonwealth University}\\
Richmond, USA\\
osmanmw@vcu.edu
}
\and
\IEEEauthorblockN{Tamer Nadeem}
\IEEEauthorblockA{\textit{Department of Computer Science} \\
\textit{Virginia Commonwealth University}\\
Richmond, USA \\
tnadeem@vcu.edu}}

\maketitle

\begin{abstract}
LoRa technology, crucial for low-power wide-area networks, faces significant performance degradation at extremely low signal-to-noise ratios (SNRs). We present LoRaFlow, a novel approach using rectified flow to reconstruct high-quality LoRa signals in challenging noise conditions. Unlike existing neural-enhanced methods focused on classification, LoRaFlow recovers the signal itself, maintaining compatibility with standard dechirp algorithms. Our method combines a hybrid neural network architecture, synthetic data generation, and robust augmentation strategies. This minimally invasive enhancement to LoRa infrastructure potentially extends operational range and reliability without overhauling existing systems. LoRaFlow opens new possibilities for robust IoT communications in harsh environments and its core methodology can be generalized to support various communication technologies.
\end{abstract}


\vspace{-0.1in}
\section{Introduction}
\vspace{-0.05in}
The rapid proliferation of the Internet of Things (IoT) has spurred the need for reliable, long-range, low-power communication technologies. IoT applications span across various domains, including smart cities, agriculture, industrial automation, and home automation, all of which demand efficient and robust communication networks \cite{adelantado2017understanding}.

Low-Power Wide-Area Networks (LPWANs) have become essential for these applications, offering extensive coverage and low power consumption. However, ensuring reliable communication in environments with low Signal-to-Noise Ratios (SNRs) is a significant challenge. This challenge is further complicated by the trade-offs between power consumption, range, and data rate inherent in LPWANs \cite{centenaro2016long,lora-alliance}.

Previous efforts to improve LoRa signal reception have included methods such as multiple gateway diversity, advanced signal processing techniques, and machine learning (ML) solutions \cite{choi2016charm, khawaja2018opr, liu2018chime, gao2018choir, deeplora, deepsense, nelora2021}. However, these approaches often necessitate dense gateway deployments, significant hardware modifications, or fail to perform optimally in highly noisy environments. 

In this paper, we introduce LoRaFlow, a novel approach that leverages advanced generative modeling techniques, specifically diffusion transformers and rectified flow, to perfectly reconstruct the original LoRa signal from noisy inputs, thereby enhancing LoRa signal demodulation. The key components of the LoRaFlow framework include a neural signal enhancement module seamlessly integrated with existing LoRa infrastructure. 

LoRaFlow offers several advantages: the ability to recover high-fidelity signals under extremely low SNR conditions, compatibility with existing LoRa systems, and minimal hardware changes. This method has the potential to extend the range and reliability of LoRa networks, thereby improving overall communication performance in IoT deployments.

The main contributions of this paper are as follows:
\begin{itemize}[left=0pt]
    \item We introduce a powerful state-of-the-art rectified flow based LoRa signal enhancement model and a novel training methodology. Our approach integrates smoothly with existing LoRa software stack and does not require removing any components unlike previous approaches. Our methodology also requires extremely minimal data collection for training. For example, we use roughly 80 times less data than previous methods\cite{nelora2021} at Spreading Factor 7.
    \item We thoroughly evaluate our approach and demonstrate almost flawless signal reconstruction in both phase and amplitude. Our signals are sufficiently perfect to defeat classification-based approaches when decoding with the simple default dechirp algorithm.
    \item We publish our complete code, model checkpoints, comprehensive evaluation results, etc. publicly under an open source license.
\end{itemize}

The remainder of this paper is organized as follows. Section~\ref{sec:background} provides an overview of LoRa signals and related work in enhancing LoRa signal reception. Section~\ref{sec:framework} details the fundamentals of diffusion models and rectified flow, which are integral to our approach, and shows an overview of the LoRaFlow framework. Sections \ref{sec:arch} and \ref{sec:training} describe the architecture and the training methodology of the LoRaFlow model, respectively. Section~\ref{sec:results} presents a comprehensive set of results demonstrating the effectiveness of our approach. Finally, Section~\ref{sec:conclusion} concludes the paper and discusses future research directions.

\vspace{-0.05in}
\section{Background \& Related Work}\label{sec:background}
\vspace{-0.05in}

\subsection{LoRa Fundamentals}\label{sec:lora_basics}
LoRa (Long Range) is a wireless technology optimized for low-power, long-distance data transmission, essential for IoT deployments in various sectors such as smart cities, agriculture, and industry. It utilizes the LoRaWAN network architecture involving end devices, gateways, network servers, and application servers, facilitating data integrity, security, and efficient routing \cite{adelantado2017understanding, centenaro2016long, lora-alliance}. Operating on sub-gigahertz unlicensed bands (433 MHz, 868 MHz, 915 MHz), LoRa employs Chirp Spread Spectrum (CSS) modulation \cite{Berni1973} that spreads the signal over a wider bandwidth. The process of chirping in LoRa involves varying the frequency of the signal over time in a linear fashion, either upwards or downwards. This technique is characterized by two key parameters: the Spreading Factor (SF) and the Bandwidth (BW).

Transmitted binary bits in LoRa are segmented into subsequences of length \( SF \), where \( SF \in [7, 12] \), forming symbols with \( M = 2^{SF} \) possible variations. The symbol rate \( R_s \) is defined as \( R_s = \frac{R_b}{SF} \), and the symbol duration \( T \) as \( T = \frac{2^{SF}}{BW} \), with BW options of 125 kHz, 250 kHz, or 500 kHz affecting data rates and signal robustness. Lower SF provides higher data rates and lower power consumption but is less reliable in terms of SNR and has a shorter range. Higher SF, on the other hand, enhances SNR reliability and range but at the cost of lower data rates and higher power consumption. Similarly, lower BW improves range and robustness but reduces data rate, while a higher BW increases data rate but decreases range and robustness.

Given the chirp rate \(k\), defined as the rate at which the chirp frequency varies over time, and formulated as \( k = \frac{BW}{T} = \frac{BW^2}{2^{SF}} \) and \(f_c\) is the carrier frequency, the form of the base chirp can be formulated as \cite{lorachirp}:

\vspace{-0.1in}
\begin{equation}
    S(t) = e^{j2\pi f_c \left(-\frac{BW}{2} + \frac{kt}{2}\right)t}, \quad t \in {\small
\left[-\frac{T}{2}, \frac{T}{2}\right]}
    \label{eqn:lorachirp}
\vspace{-0.05in}
\end{equation}

\begin{figure*}[th!]
    \centering
    \includegraphics[width=0.9\textwidth,keepaspectratio]{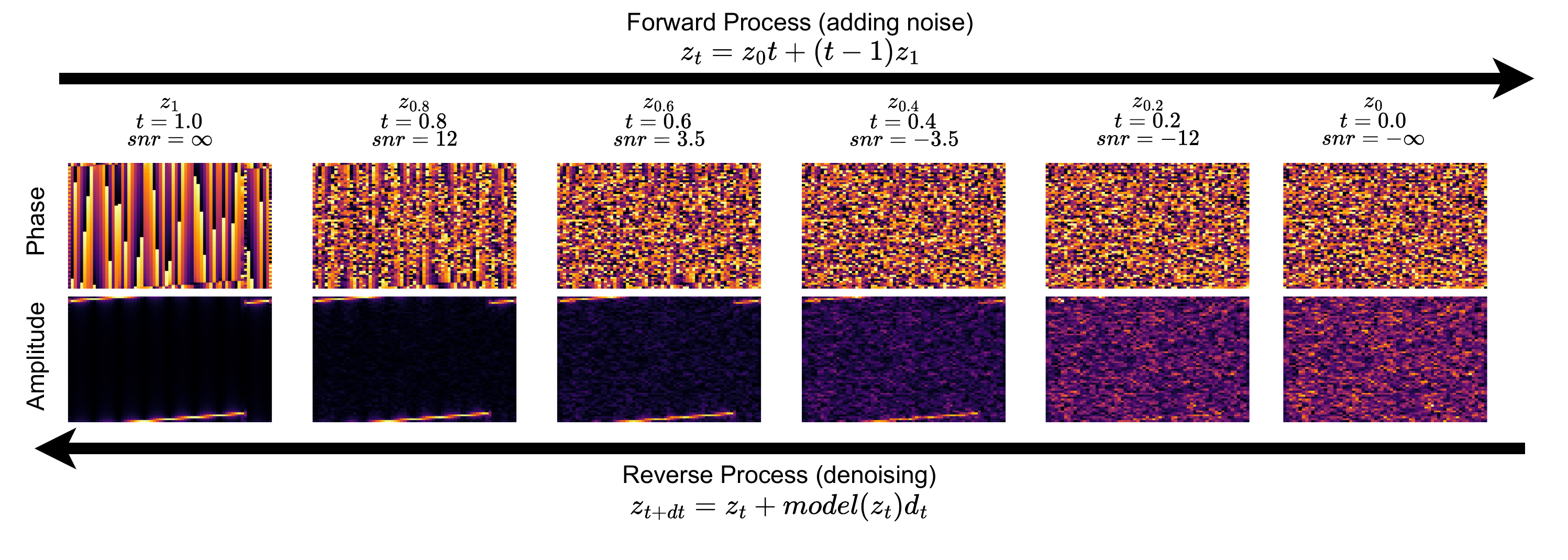}
    \vspace{-0.2in}
    \caption{Overview of the iterative refinement process using diffusion models for LoRa signals. The forward process adds noise (defined in closed-form), while the reverse process denoises the signal using our model. Signals are presented as STFTs where the top row represents phase while the bottom row represents amplitude.}
    \label{fig:iterative_refinement}
    \vspace{-0.2in}
\end{figure*}

LoRa modulation uniquely maps each of the \( M \) symbols to a distinct chirp. The chirp for the \( m^{th} \) symbol is generated by time-shifting the base chirp by \(\tau_m=\frac{m}{BW}\), where \(\tau_m = \frac{i}{2^{SF}} \cdot T\). Any part of the chirp outside \(\left[-\frac{T}{2}, \frac{T}{2}\right]\) is cyclically adjusted back into the interval. Figure \ref{fig:iterative_refinement} provides a visual example of this modulation process.

At the receiver, the chirp signal is dechirped by mixing it with a reference chirp, converting frequency variations into a baseband signal for demodulation. The process includes multiplying by the conjugate of the reference chirp and using Fast Fourier Transform (FFT) to detect peak frequencies of transmitted symbols. In low SNR conditions, signal quality degrades, particularly over long distances due to path loss, challenging reliable communication. Efficiently reconstructing the noisy received signal is essential for accurate dechirping and decoding, crucial for energy conservation in low-power IoT devices \cite{centenaro2016long, lora-alliance, adelantado2017understanding}.

\subsection{Related Work}
Recent efforts to enhance the reception of LoRa signals under low SNR conditions have focused on leveraging multiple gateways, hardware diversity, and advanced signal processing techniques. Four notable approaches are Charm, OPR, Chime, and Choir. Charm improves energy efficiency by using the spatial diversity of multiple gateways to decode weak chirp symbols through coherent combining, achieving 1-3 dB SNR gain with 2-8 gateways per node \cite{choi2016charm}. OPR explores disjoint link-layer bit errors across multiple gateways to recover corrupted packets, resulting in 1.5-2.5 dB SNR gain with 2-6 gateways per node \cite{khawaja2018opr}. Chime utilizes heartbeat packets and three gateways to estimate wireless channel states and select optimal frequencies for transmission, achieving 2.4-3.4 dB SNR gain with 4-6 gateways per node \cite{liu2018chime}. Choir leverages hardware diversity with up to 36 co-located LoRa nodes to boost received signal strength but does not provide specific SNR gain values \cite{gao2018choir}. These approaches, however, require dense deployment of gateways and nodes, which may not be cost-effective or feasible in all scenarios.

While recent advancements in machine learning (ML) have opened new avenues for signal procssing, very few works considered LoRa signals. 
DeepLoRa \cite{deeplora} employs a Bi-LSTM DNN to create a land-cover-aware path loss model, reducing estimation error to less than 4 dB. DeepSense \cite{deepsense} extends this approach, exploring deep learning-augmented random access for LPWAN coexistence, even under extreme noise conditions (e.g., -10 dB). A very recent work NELoRa \cite{nelora2021} introduces a neural-enhanced demodulation method that exploits deep learning to support ultra-low SNR LoRa communication. The architecture of NELoRa comprises two key components: a mask-enabled Deep Neural Network (DNN) filter and a spectrogram-based DNN decoder. The filter aims to recover clean chirp symbols by masking their noisy input spectrogram, effectively separating the signal from noise. Subsequently, the decoder classifies the recovered chirp symbols, exploiting the finite coding space of LoRa to its advantage. This method achieves SNR gains of 1.84-2.35 dB, outperforming traditional dechirp methods and other ML-based approaches by lowering the SNR threshold for chirp symbol decoding.

Our work builds upon these foundations, taking a fundamentally different approach. Instead of focusing on classification, we aim to recover the LoRa signal itself at extremely low SNRs using rectified flow, a diffusion-based generative modeling technique. This signal recovery approach offers the potential to operate at even lower SNRs than NELoRa, while maintaining compatibility with existing dechirp algorithms. By reconstructing the signal prior to classification, our method opens up new possibilities for pushing the boundaries of LoRa communication in challenging environments.

\section{LoRaFlow Framework Overview}\label{sec:framework}
In this section, we present the LoRaFlow framework starting by discussing the core approach of LoRaFlow, followed by an in-depth look at the fundamental diffusion models and rectified flow mechanisms that underpin its operation. Finally, we illustrate the practical implementation of LoRaFlow within the current LoRa network architecture.

\subsection{LoRaFlow Approach}
Enhancing signal reception under extremely low Signal-to-Noise Ratio (SNR) conditions remains a significant challenge in LoRa communication. Prior works primarily focus on classifying received noisy signals into corresponding symbols. Although effective to some extent, these methods often struggle in scenarios with severe signal degradation, limiting the reliability and range of LoRa networks. Our LoRaFlow framework addresses this issue by reconstructing the received noisy LoRa signal into a full, denoised signal at the LoRa gateways. Complete reconstruction is a substantially harder task than classification. By achieving strong signal reconstruction, our method can significantly enhances communication reliability under extreme noise conditions.

Reconstructing the entire transmitted signal significantly enhances communication systems, improving accuracy, security, and overall performance for a range of applications. This comprehensive approach ensures precise data representation, reduces transmission errors, and bolsters signal integrity even in adverse conditions. Full signal access supports robust security protocols, including advanced authentication and detailed signal fingerprinting, while also extending operational ranges and improving data recovery in challenging environments. Sophisticated signal processing and decoding techniques are enabled, optimizing network resource utilization and supporting high-reliability applications such as remote sensing and autonomous vehicles. Additionally, this method enhances the system's resilience to sophisticated attacks like spoofing and replay, facilitating accurate transmitter identification and enabling precise forensic analysis in security breach scenarios. LoRaFlow approach in full signal reconstruction pushes the boundaries of LoRa communication, offering a significant advancement in the field and opening new avenues for research and application in low-power, long-range communication technologies.

\subsection{Fundamentals of LoRaFlow: Diffusion Models and Rectified Flow}
Achieving high-fidelity signal recovery at extremely low SNRs necessitates the use of sophisticated generative models capable of reconstructing signals buried in noise. Diffusion-based generative models have recently emerged as powerful tools for synthesizing complex data distributions. These models operate by progressively corrupting data with noise and then learning to reverse this process to generate samples by iteratively denoising random noise.

The core principle behind this reversal process is the \emph{score function}, the gradient of the log probability density with respect to the data. Knowing the score function allows the model to move from regions of low probability (high noise) towards regions of high probability (low noise). Early diffusion models, such as Noise Conditional Score Networks (NCSN) \cite{song2019generative} and Denoising Diffusion Probabilistic Models (DDPM) \cite{ho2020denoising}, introduced methods for learning these score functions at discrete noise levels.

Song et al. \cite{song2021scorebasedgenerativemodelingstochastic} generalized this approach using Stochastic Differential Equations (SDEs), which allow for continuous-time diffusion processes. The forward SDE is given by:

\vspace{-0.15in}
\begin{equation}
\label{eqn:sde_lora}
\small
d \mathbf{x}_t = \mathbf{f}(\mathbf{x}_t, t) dt + g(t) d\mathbf{w}_t,
\vspace{-0.1in}
\end{equation}

where \(\mathbf{x}_t\) represents the signal at time \(t\), \(\mathbf{f}(\mathbf{x}_t, t)\) is the drift coefficient, \(g(t)\) is the diffusion coefficient, and \(d\mathbf{w}_t\) represents an infinitesimal increment of a standard Wiener process. The corresponding reverse-time SDE is:

\vspace{-0.2in}
\begin{equation}
\label{eqn:reverse-sde_lora}
\small
d\mathbf{x}_t = [\mathbf{f}(\mathbf{x}_t, t) - g(t)^2 \nabla_{\mathbf{x}} \log p_t(\mathbf{x}_t) ]dt + g(t) d\bar{\mathbf{w}}_t,
\vspace{-0.1in}
\end{equation}

where \(\bar{\mathbf{w}}_t\) is a Wiener process in reverse time, and \(p_t(\mathbf{x}_t)\) denotes the probability density of \(\mathbf{x}_t\). This reverse-time SDE depends on the score function and is crucial for the generative process. To address the computational demands of solving these equations, the concept of a \emph{Probability Flow (PF) Ordinary Differential Equation (ODE)} was introduced. The PF ODE provides a deterministic counterpart to the stochastic diffusion process, allowing for more efficient sample generation. The PF ODE is given by:

\vspace{-0.15in}
\begin{equation}
\label{eqn:prob-flow_lora}
\small
\frac{d\mathbf{x}_t}{dt} =  \mathbf{f}(\mathbf{x}_t, t) - \frac{1}{2}g(t)^2 \nabla_{\mathbf{x}} \log p_t(\mathbf{x}_t),
\vspace{-0.1in}
\end{equation}

Rectified flow \cite{liu2022flow, liu2022rectified} further improves this by learning a mapping with \emph{straight-line} trajectories. This is achieved by minimizing the objective:

\vspace{-0.15in}
\begin{equation}
\label{eqn:rectified-flow_lora}
\small
\min_v \int_0^1 \mathbb{E} \left[ \|\mathbf{X}_1 - \mathbf{X}_0 - v(t \mathbf{X}_1 + (1-t)\mathbf{X}_0, t)\|_2^2 \right] dt,
\vspace{-0.05in}
\end{equation}

where \((\mathbf{X}_0, \mathbf{X}_1)\) is a coupling of the noise and data distributions, and \(v\) is the velocity field of the ODE. This approach directly maps noise to data, reducing the number of steps needed for high-fidelity signal recovery, making it particularly suitable for low SNR conditions.

Architectural choices are crucial for the performance of diffusion models. While U-Net architectures have been widely used due to their ability to process spatial information efficiently at multiple scales, transformer-based architectures like Diffusion Transformers (DiT) introduced by Peebles and Xie \cite{peebles2023scalable} have shown remarkable scalability and efficiency. DiT leverages the self-attention mechanism to capture long-range dependencies, making it highly suitable for high-dimensional data processing.

Given the strengths of both U-Net and transformer-based architectures, hybrid approaches combining elements of both designs offer significant advantages. Our approach integrates DiT blocks with convolutional down/upsampling layers. This hybrid architecture leverages the scalability and flexibility of transformers while retaining the U-Net's ability to process multi-scale spatial information effectively. Further, it allows us to partially defeat the Transformer's quadratic compute complexity along the sequence length. By exploring this hybrid space, we aim to achieve superior performance in handling complex signal recovery tasks with improved quality and computational efficiency.

An example of our iterative refinement process for LoRa signals is illustrated in Figure \ref{fig:iterative_refinement}. The figure shows the forward process (top row) and the reverse process (bottom row). In the forward process, noise is progressively added to the signal, transitioning from a high SNR to a low SNR state. This illustrates how the original signal is corrupted over time. The reverse process then denoises the signal, iteratively refining it back to a high-fidelity state. This illustrates how the model reconstructs the original signal from noisy data. 

During inference, the reverse process is discretized into N steps marked by unique t values, each of which requires a neural function evaluation (NFE) which is one forward-propagation of the model in our case. 

To map SNRs to the time parameter t in our model, we use the following function:

\vspace{-0.1in}
\begin{equation}
\small
t = \frac{\sqrt{SNR}}{1 + \sqrt{SNR}}, \quad \text{where } SNR = 10^{SNR_{dB}/10}
\vspace{-0.05in}
\end{equation}

This mapping allows us to associate different noise levels with specific points in the ODE trajectory. In our inference scenario, we're able to skip an appropriate amount of noise steps and allows us to "insert" the received samples at the appropriate place in the ODE. This in turn leads to a natural property where the less noisy a signal is, the faster our model executes and vice versa.

Our LoRaFlow framework utilizes these advanced generative modeling techniques of diffusion models, the efficiency of rectified flow, and the robustness of hybrid DiT architectures to achieve high-fidelity recovery of LoRa signals even when significantly corrupted by noise.

\begin{figure*}[t!]
    \centering
    \includegraphics[width=0.9\textwidth,keepaspectratio]{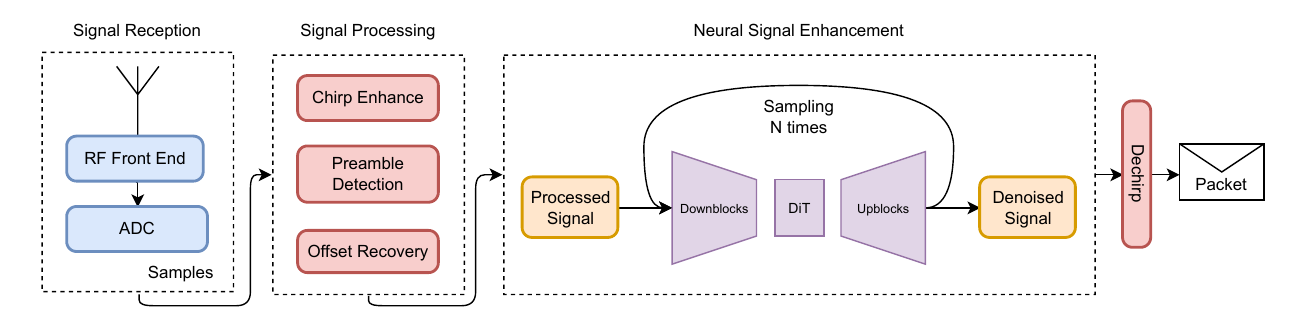}
    \vspace{-0.1in}
    \caption{Integration of LoRaFlow at LoRa gateways.}
    \label{fig:loraflow_framework}
    \vspace{-0.2in}
\end{figure*}

\subsection{LoRaFlow in Practice}
Designed to complement existing LoRa infrastructure rather than replace it, LoRaFlow takes a minimally invasive approach. Unlike more radical redesigns such as NELoRa \cite{li2021nelora}, which overhaul significant portions of the LoRa demodulation pipeline, our method focuses on a single, critical intervention: denoising the raw signal before it reaches the standard dechirp operation. This allows for seamless integration into current LoRa networks without substantial modifications.

The integration of our method into the LoRa framework can be summarized into three main components (see Figure \ref{fig:loraflow_framework}):
\begin{enumerate}[left=0pt]
    \item Signal Reception and Initial Processing: Similar to the NELoRa approach, after the initial signal reception by the RF front end and conversion to digital samples by the ADC, the signal undergoes traditional processing steps. These include chirp enhancement, preamble detection, and offset recovery. This stage prepares the signal for the subsequent denoising process by correcting any initial distortions and aligning the signal for optimal processing.

    \item LoRaFlow Signal Denoising: This is the core innovation of our approach. The processed signal then enters our neural signal enhancement module. Here, the signal is denoised up to \( N \) times (depending on the SNR as we skip a number of steps as mentioned previously) using our model. This module employs rectified flow—a diffusion-like generative modeling technique—to effectively denoise the signal. This denoising process is crucial for separating the LoRa chirps from background noise, thereby improving the quality of the signal before it undergoes standard demodulation.

    \item Standard Demodulation: The denoised signal proceeds to the standard LoRa demodulation pipeline, starting with the dechirp operation. After this stage, the denoised signal is transformed into packets for further application processing. It is notable that our approach is entirely orthogonal to classification-based approaches. The combination of both while not complicated to implement is left to future work. 
\end{enumerate}

Mathematically, we can express the integration of our method as a preprocessing step:

\vspace{-0.15in}
\[
s_{\text{denoised}}(t) = \mathcal{F}(s_{\text{received}}(t))
\vspace{-0.05in}
\]

where \( s_{\text{received}}(t) \) is the raw received signal, \( \mathcal{F} \) represents our rectified flow denoising operation, and \( s_{\text{denoised}}(t) \) is the cleaned signal that is then fed into the standard dechirp process:

\vspace{-0.25in}
\[
s_{\text{dechirped}}(t) = s_{\text{denoised}}(t) \cdot s_{\text{base}}^*(t)
\vspace{-0.05in}
\]

Here, \( s_{\text{base}}^*(t) \) is the complex conjugate of the base chirp signal.

Our approach offers several distinct advantages. By preserving the dechirp operation and subsequent processing steps, LoRaFlow maintains full \textbf{compatibility} with existing LoRa hardware and software. The integration requires minimal changes to current LoRa systems, which can accelerate adoption and deployment, showcasing its \textbf{simplicity}. Additionally, the denoising step is mostly or entirely bypassed in high SNR conditions, allowing for adaptive use based on signal quality, demonstrating its \textbf{flexibility}.

Focusing on signal recovery at extremely low SNRs, our method enhances the strengths of LoRa modulation rather than replacing them. This conservative enhancement approach allows for incremental improvements to LoRa systems, potentially extending their operational range and reliability without overhauling existing infrastructure. The simplicity and effectiveness of our integration strategy position LoRaFlow as a practical advancement in LoRa communication, bridging the gap between cutting-edge machine learning techniques and the real-world constraints of IoT deployment. As LoRa remains essential for long-range, low-power communication, our approach pushes the boundaries of its capabilities while maintaining the robustness and reliability that have made it a cornerstone of IoT networks.

\section{LoRaFlow Model Architecture} \label{sec:arch}
Our proposed architecture for the LoRaFlow denoising model integrates elements from advanced diffusion transformers \cite{katherine_crowson_2023_10284390} with domain-specific adaptations tailored for LoRa signal processing. The model is composed of three main components: an input processing stage, a transformer core, and an output stage. This setup is then enhanced with an auxiliary classifier during training. These components are tied to the "Neural Signal Enhancement" module in Figure \ref{fig:loraflow_framework}, which showcases our method.

\begin{figure}[htbp]
    \vspace{-0.2in}
    \centering
    \includegraphics[width=0.8\columnwidth]{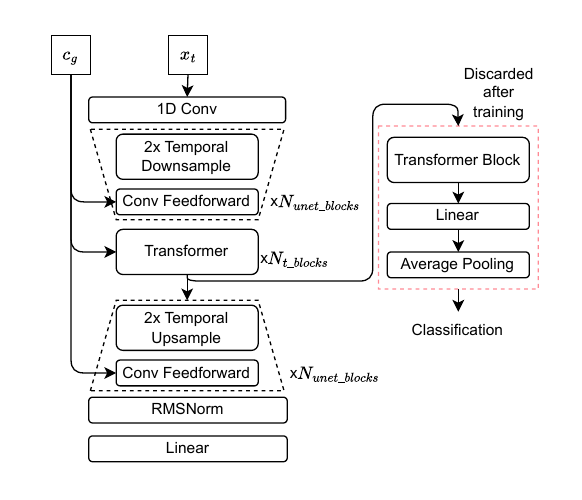}
    \vspace{-0.2in}
    \caption{LoRaFlow Model Architecture.}
    \label{fig:loraflow_main_design}
    \vspace{-0.25in}
\end{figure}

The input processing stage employs a series of convolutional layers and temporal downsampling modules that progressively downsample the input signal while increasing the feature dimension. This stage serves two primary purposes: it extracts low-level features from the raw signal and reduces the sequence length, making subsequent transformer operations more computationally efficient. Specifically, this stage includes a single 1D convolutional layer followed by 2x temporal downsampling and convolutional feedforward layers, which efficiently compress the input signal's temporal dimensions.

\begin{figure*}[hbtp!]
    \centering
    \includegraphics[width=0.8\textwidth]{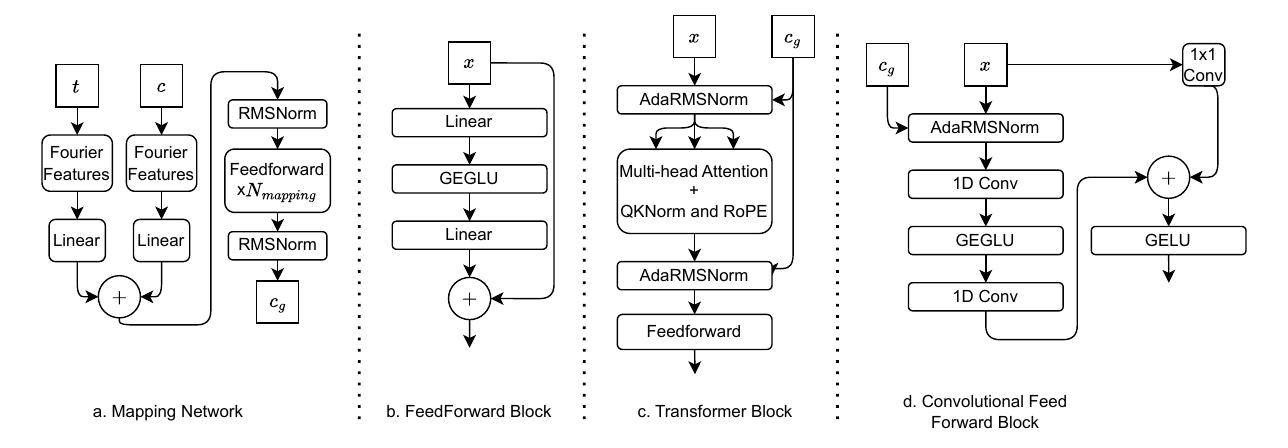}
    \vspace{-0.15in}
    \caption{Detailed Components of the LoRaFlow Architecture}
    \vspace{-0.2in}
 \label{fig:loraflow_components_details}
\end{figure*}

The core of our model is a transformer block adapted from the Diffusion Transformer (DiT) architecture.\cite{peebles2023scalable, vaswani2017attention} This block processes the latent representation produced by the input stage, capturing long-range dependencies and global context critical for understanding LoRa chirp structures. The self-attention mechanism within the transformer is particularly well-suited for modeling the phase relationships in LoRa signals across different time scales. The DiT architecture includes several key components shown in Figure \ref{fig:loraflow_components_details}. The \emph{mapping network} generates conditioning embeddings from the time step \( t \) and conditional embeddings \( c \). Fourier features are applied to these inputs to capture these scalar inputs as accurately as possible, followed by linear layers and RMS normalization (RMSNorm). This results in the generation of conditioning embeddings \( c_g \) (see Figure \ref{fig:loraflow_components_details}a). The \emph{feedforward block} consists of a series of linear layers and Gated Linear Units (GELU)\cite{hendrycks2016gaussian} activations. The GELU activation function helps to introduce non-linearity, which is crucial for capturing complex signal relationships. The \emph{feedforward block} is responsible for processing the intermediate representations within the transformer (see Figure \ref{fig:loraflow_components_details}b). At the heart of the DiT architecture, the \emph{transformer block} includes multi-head attention mechanisms, which allow the model to attend to different parts of the input sequence simultaneously. This mechanism is augmented with QKNorm (Query-Key Normalization) and RoPE (Relative Position Encodings) to better capture positional information. The block also includes AdaRMSNorm (adaptive RMS normalization) layers and feedforward networks (see Figure \ref{fig:loraflow_components_details}c) as is standard in Transformers. The \emph{convolutional feedforward block} integrates convolutional layers with the same layout as the feedforward blocks networks. 1D convolutions are used to capture local dependencies in the signal. AdaRMSNorm is applied to maintain stable training dynamics and condition the network (see Figure \ref{fig:loraflow_components_details}d).

Following the transformer core, the output stage mirrors the input stage in reverse, using convolutional layers and temporal upsampling modules to upsample the signal back to its original dimensions. This stage reconstructs the denoised signal from the processed latent representation. The careful design of the upsampling modules ensures that the high-resolution features are accurately restored, maintaining the integrity of the denoised signal. This stage specifically includes 2x temporal upsampling and convolutional feedforward layers, with RMSNorm and linear layers to refine the signal.

A key innovation in our architecture in Figure \ref{fig:loraflow_main_design} is the incorporation of an auxiliary classifier connected to the transformer's output. This classifier is designed to predict the LoRa chirp class (which is unique per spreading factor) from the latent representation. Importantly, the classifier is only used during training and is discarded during inference. This approach guides the model to learn more discriminative features without constraining its generative capabilities, enhancing the overall performance of the denoising process.

The model processes time embeddings and augmentation condition embeddings, which are injected into each block of the network. These embeddings allow the model to adapt its behavior based on the noise level and specific augmentations applied to the input signal. The incorporation of these embeddings ensures that the model remains flexible and responsive to varying signal conditions, a crucial feature for effective denoising in low SNR environments. It should be noted that \(x_t\) shown in Fig \ref{fig:loraflow_main_design} refers to a noisy sample at timestep t while \(c_g\) refers to the output of the Mapping Network.

\section{LoRaFlow Model Training Methodology}\label{sec:training}
In this section, we detail the comprehensive training methodology employed to optimize the LoRaFlow model. This methodology includes the design of multi-component loss functions, synthetic data generation, and data augmentation techniques, ensuring the model's generalizability to real-world data.

\subsection{Loss Functions and Training Objectives}
Our training process employs a multi-component loss function designed to address the unique challenges of LoRa signal reconstruction at extremely low SNRs. The total loss is a weighted sum of four components:

\vspace{-0.15in}
\begin{equation}
\small
\mathcal{L}_{\text{total}} = \mathcal{L}_{\text{recon}} + \lambda_1 \mathcal{L}_{\text{FFT}} + \lambda_2 \mathcal{L}_{\text{STFT}} + \lambda_3 \mathcal{L}_{\text{cls}}
\vspace{-0.1in}
\end{equation}

where $\lambda_1$, $\lambda_2$, and $\lambda_3$ are weighting coefficients.

The primary reconstruction loss, $\mathcal{L}_{\text{recon}}$, is based on the rectified flow formulation:

\vspace{-0.2in}
\begin{equation}
\small
\mathcal{L}_{\text{recon}} = \mathbb{E}_{t \sim U(0,1)} \left[ \|z_1 - z_0 - v_\theta(t z_1 + (1-t)z_0, t)\|_2^2 \right]
\vspace{-0.1in}
\end{equation}

where $z_0$ is the noisy input, $z_1$ is the clean target, and $v_\theta$ is our model.

To enhance frequency-domain fidelity, we incorporate two spectral losses. The FFT loss, $\mathcal{L}_{\text{FFT}}$, computes the discrepancy between the Fourier transforms of the predicted and target signals after applying the LoRa chirp:

\vspace{-0.1in}
\begin{equation}
\small
\mathcal{L}_{\text{FFT}} = \text{HuberLoss}(\text{FFT}(x_{\text{pred}} \cdot c), \text{FFT}(x_{\text{target}} \cdot c))
\vspace{-0.05in}
\end{equation}

where $c$ is the LoRa chirp and $x$ represents the complex-valued signals.

The multi-scale STFT loss, $\mathcal{L}_{\text{STFT}}$, captures time-frequency characteristics at various resolutions:

\vspace{-0.15in}
\begin{equation}
\small
\mathcal{L}_{\text{STFT}} = \sum_{(n,h) \in S} \text{HuberLoss}(\text{STFT}_n^h(x_{\text{pred}}), \text{STFT}_n^h(x_{\text{target}}))
\vspace{-0.05in}
\end{equation}

where $S$ is a set of (window size, hop length) pairs.

The classification loss, $\mathcal{L}_{\text{cls}}$, is a combination of cross-entropy and an auxiliary regularization term:

\vspace{-0.15in}
\begin{equation}
\small
\mathcal{L}_{\text{cls}} = \text{CrossEntropy}(y_{\text{pred}}, y_{\text{true}}) + \alpha \|(\log \sum_i \exp(y_{\text{pred},i}))\|_2^2
\vspace{-0.1in}
\end{equation}

where $\alpha$ is a small coefficient (e.g., 1e-4) to keep the logits normalized \cite{chowdhery2023palm}.

This classification loss serves a crucial role in our training process. The similarity between different LoRa chirps, which can be viewed as rotations of each other in the signal space, poses a challenge for reconstruction-based losses alone. These losses may not provide sufficient guidance for the model to distinguish between similar chirps accurately. By introducing the classification loss, we encourage the model's latent representations to be more discriminative, helping it to "choose correctly" among similar chirp candidates during reconstruction.

The auxiliary classifier is designed with additional parameters to contain potential representation collapse often associated with classification losses. These extra parameters are removed during inference, ensuring that the final model retains the flexibility needed for high-quality signal reconstruction while benefiting from the improved feature learning during training.

This multi-component loss function, combined with our hybrid architecture, enables our model to achieve high-fidelity LoRa signal recovery at extremely low SNRs, effectively pushing the boundaries of reliable communication in challenging environments.

\begin{figure}[tbp]
    \centering
    \includegraphics[width=0.75\columnwidth]{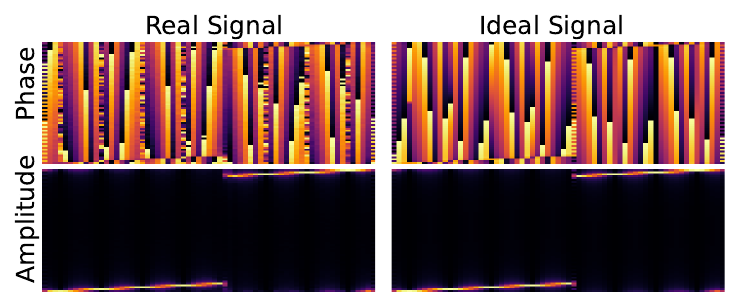}
     \vspace{-0.1in}
    \caption{Showcase of a sample from the NELoRa dataset compared to our dataset of ideal synthesized data.}
    \label{fig:comparison_real_vs_ideal}
    \vspace{-0.2in}
\end{figure}
\subsection{Synthetic Data Generation for Training}
To train our model effectively, we developed a comprehensive synthetic LoRa signal dataset that captures the diversity of real-world LoRa transmissions. Our dataset - taken from NELoRa - encompasses a wide range of spreading factors (SF) specifically: 7, 8, 9, and 10. For each spreading factor choice, we generate chirp signals as described earlier using Equation \ref{eqn:lorachirp}. We generate both upchirps (increasing frequency) and downchirps (decreasing frequency) to represent the full range of LoRa symbols. To encode different symbols, we apply appropriate time shifts \( \tau_m \) as described in Subsection \ref{sec:lora_basics}. This approach allows us to generate all possible symbols for each SF, resulting in a dataset that comprehensively covers the LoRa signal space. We show an example of a real data sample compared to a synthetic sample in Figure \ref{fig:comparison_real_vs_ideal}

The use of synthetic data offers several advantages. First, it allows us to generate a large, diverse dataset without the need for extensive real-world data collection. Second, it provides perfect ground truth for training, free from real-world channel impairments. Finally, it enables us to systematically explore the full range of LoRa configurations, ensuring our model's generalizability.

\subsection{Data Augmentation Techniques}

To enhance our model's robustness and generalization capabilities, we implement a suite of data augmentation techniques tailored to LoRa signals and potential channel effects. These augmentations are applied probabilistically during training, each with a base probability of 0.15. We use an 8-dimensional input condition vector \(c\), where 7 dimensions correspond to on/off augmentations (1 or -1), and the 8th denotes the current SF. This vector is dropped out to all 0s with a 10\% probability, challenging the model to infer augmentations and SF independently. All evaluations are performed with \(c\) set to 0s, which we find to be highly effective. While we employ several strategies, we focus on describing our frequency-domain masking technique in detail due to its crucial role in simulating real-world channel impairments.

\subsubsection{Frequency-Domain Masking}
Frequency-domain masking is particularly relevant for LoRa signals, as it simulates frequency-selective fading and interference, which are common challenges in wireless communications. This augmentation operates on the Short-Time Fourier Transform (STFT) representation of the signal, allowing us to manipulate its time-frequency characteristics.

The process of frequency-domain masking can be summarized in Algorithm \ref{alg:freq_domain_masking}.

\vspace{-0.1in}
\begin{lstlisting}[language=Python, caption={Frequency-Domain Masking},label={alg:freq_domain_masking}]
def frequency_domain_masking(signal):
    stft = compute_stft(signal)
    max_num_masks, max_mask_size = 2, 2
    
    for dim in ['time', 'frequency']:
        num_masks = random_int(1, max_num_masks)
        for _ in range(num_masks):
            mask_size = random_int(1, max_mask_size)
            mask_start = random_int(0, stft.shape[dim] - mask_size)
            mask_value = random_uniform(0, 0.5) if random() < 0.5 else 0
            apply_mask(stft, dim, mask_start, mask_size, mask_value)
    
    return compute_istft(stft)
\end{lstlisting}

This augmentation technique randomly attenuates or completely masks small regions of the spectrogram. The process is applied independently to both time and frequency dimensions, allowing for a diverse range of potential distortions. The number of masks applied in each dimension is randomly chosen between 1 and 2, with each mask covering 1 or 2 time/frequency bins. The attenuation value is either 0 (complete masking) or a random value between 0 and 0.5 (partial attenuation). Masks are applied to the real and imaginary components of the STFT independently, allowing for phase distortions as well as amplitude changes. By applying these masks, we simulate various channel effects such as narrow-band interference, frequency nulls, and short-term fading, thereby improving our model's ability to handle these phenomena in real-world scenarios.

\subsubsection{Additional Augmentation Techniques}

In addition to frequency-domain masking, we employ several other augmentation strategies:

\begin{itemize}
    \item \textbf{Time-domain shifts}: We randomly roll the signal in the time domain, simulating the effect of symbol boundary misalignment.
    \item \textbf{Signal inversion}: The entire signal is inverted with a certain probability, helping the model become invariant to phase changes that might occur during transmission.
    \item \textbf{Spectrogram rolling}: We implement rolling operations on the spectrogram in both time and frequency dimensions, simulating minor frequency offsets and time shifts.
\end{itemize}

The full implementation details of these augmentations, including their probabilistic application, are available in our open-source code release.\footnote{DOUBLE BLIND}

By combining these diverse augmentation techniques, we create a robust training regime that prepares our model for the challenges of real-world LoRa signal recovery at extremely low SNRs. This approach significantly enhances the model's ability to generalize across a wide range of signal conditions and channel impairments, ultimately improving its performance in practical deployments.

\begin{figure*}[hbtp!]
    \centering
    \includegraphics[width=0.9\textwidth]{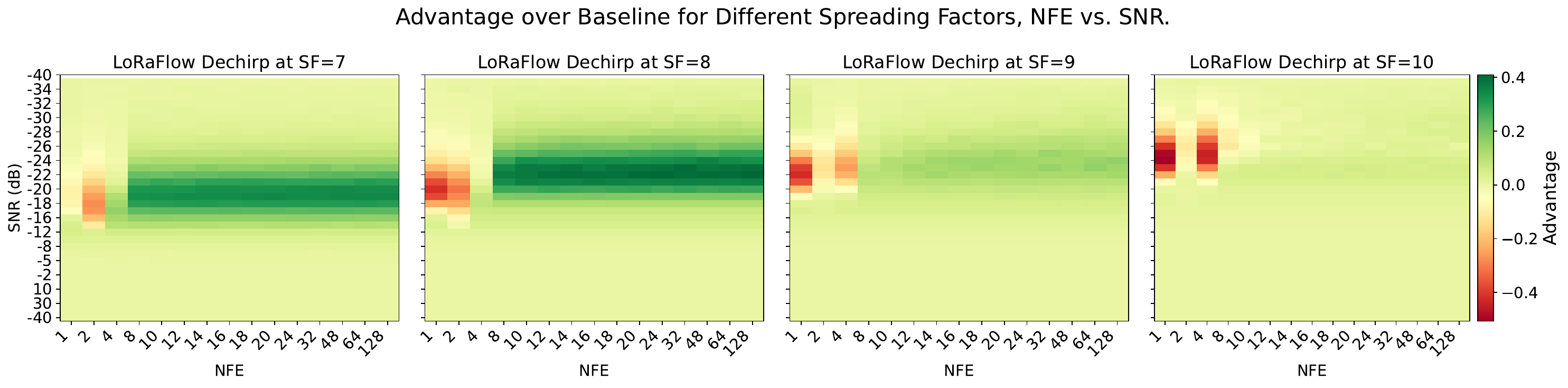}
    \vspace{-0.1in}
    \caption{This figure shows the accuracy advantage over baseline dechirp for all SNRs tested and various numbers of neural function evaluations (NFE). Light green/yellow indicates equal performance to baseline, red indicates a performance degradation, and green indicates performance improvement.}
    \label{fig:advantage_plot_nfe}
    \vspace{-0.2in}
\end{figure*}

\subsection{Generalization to Real-World Data}
While our approach demonstrates impressive performance on synthetic data, bridging the gap to real-world LoRa signals is crucial for practical deployment. To address this challenge, we employ a fine-tuning strategy that leverages the NELoRa dataset \cite{li2021nelora}, adapting our model to the nuances of actual LoRa transmissions while maintaining its ability to operate at extremely low SNRs. The NELoRa dataset comprises 27,329 LoRa symbols covering spreading factors from 7 to 10, collected in real-world indoor environments.

Our fine-tuning process is designed to be data-efficient, recognizing the scarcity of labeled real-world LoRa signals, especially at very low SNRs. We adopt a one-shot learning scenario, utilizing a single sample per class from the NELoRa dataset. This approach minimizes the need for extensive real-world data collection and demonstrates the robustness of our model architecture.

The fine-tuning procedure closely mirrors our initial training process:
\begin{math}
\small
\mathcal{L}_{\text{fine-tune}} = \mathbb{E}_{z_0, z_1 \sim p_{\text{data}}} \left[ \int_0^1 \mathbb{E}_t \left[ \|z_1 - z_0 - v_\theta(tz_1 + (1-t)z_0, t)\|_2^2 \right] dt \right]
\end{math}

where $p_{\text{data}}$ now represents a mixture of real and synthetic data distributions. Specifically, during training we take samples from real NELoRa dataset with 95\% probability and from our synthetic dataset with 5\% probability. Do note that in either case we only have one sample available per class. This mixture strategy serves two purposes: it prevents catastrophic forgetting of the rich patterns learned from synthetic data and fills the gaps where the NELoRa dataset sometimes has only 1 sample for some classes, in which case, we leave the class out and rely entirely on synthetic data.

The use of the same loss function and optimization setup as in the initial training phase allows for a seamless transition between synthetic and real data. This continuity in the learning process facilitates efficient knowledge transfer, enabling the model to quickly adapt to the characteristics of real LoRa signals while retaining its ability to operate in extremely challenging noise conditions.

Empirically, we observe that this fine-tuning approach leads to significant improvements in real-world performance, particularly in scenarios where the SNR is well below the theoretical limits of traditional LoRa demodulation techniques. The model's ability to generalize from a single real-world example per class underscores the effectiveness of our rectified flow-based approach in capturing the fundamental structure of LoRa signals, transcending the specifics of synthetic data generation.

\section{Performance Evaluation}\label{sec:results}

We rigorously evaluate LoRaFlow's performance across various dimensions to demonstrate its efficacy in enhancing LoRa signal reception under challenging low SNR conditions. Our evaluation metrics include accuracy advantage over baseline dechirp, qualitative signal reconstruction assessment, and quantitative comparison with the state-of-the-art NELoRa method using Signal Error Rate (SER) advantage. We also utilize the Area Under Curve (AUC) metric to provide a holistic performance summary across all SNRs.

\subsection{Experimental Setup}

Our training process leverages synthetic data with a single sample per class, necessitating a robust augmentation strategy. We employ dynamic batch sizes tailored to each Spreading Factor (SF): 2048 for SF7, 1024 for SF8, 512 for SF9, and 256 for SF10. For the scope of this work, we use a bandwidth of 125,000 for all experiments. The model undergoes 300,000 updates on synthetic data, followed by fine-tuning with real data. For fine-tuning, we select one example per class from the NELoRa dataset, training on real data for 95\% of the subsequent 50,000 updates and on synthetic data for the remainder. Despite one-shot (1 sample per class) finetuning being the most difficult adaptation scenario, we find that our model excels in it easily.

Training utilizes 6 NVIDIA SXM5 H100 GPUs, optimized with AdamW Schedule Free \cite{defazio2024road}. We implement Flash Attention \cite{dao2022flashattention} and leverage \texttt{torch.compile} with dynamic sizes to enhance computational efficiency. To ensure reproducibility, we use a fixed random seed and deterministic splitting across all experiments.

For evaluation, we assess performance across a range of SNRs (-40dB to -10dB) and compare against both traditional dechirp methods and the state-of-the-art NELoRa approach. Our metrics include accuracy advantage, Signal Error Rate (SER) advantage, and Area Under Curve (AUC) for comprehensive performance assessment.

\subsection{Impact of Neural Function Evaluations on Performance}

Figure \ref{fig:advantage_plot_nfe} shows the accuracy advantage of LoRaFlow over the baseline dechirp method across various Spreading Factors (SFs) and Signal-to-Noise Ratios (SNRs), plotted as a function of the number of Neural Function Evaluations (NFE). This comprehensive analysis reveals several key insights. LoRaFlow demonstrates substantial performance improvements under certain scenarios, particularly for SFs 7 and 8 within the mid-range SNRs (-30 to -20 dB), where accuracy improvements of up to 0.4 are observed. The influence of NFE is significant, as increases in NFE lead to enhanced performance. However, increasing NFE beyond 16 does not seem to yield significant benefits, suggesting that the model's trajectory is simple enough to be sufficiently well approximated by that level of discretization. While improvements for SFs 9 and 10 are less dramatic, they remain substantial, particularly between -25 and -15 dB SNRs. At extremely low SNRs (below -35 dB), the performance of LoRaFlow aligns with the baseline, indicating a potential limitation in signal recovery under severe noise conditions. These findings emphasize LoRaFlow's capacity to enhance LoRa communication reliability in noisy environments, particularly at lower SFs where other methods falter. The results also highlight the crucial role of balancing computational load against signal recovery effectiveness in real-world applications.

\begin{figure*}[hbtp!]
    \centering
    \includegraphics[width=0.8\textwidth]{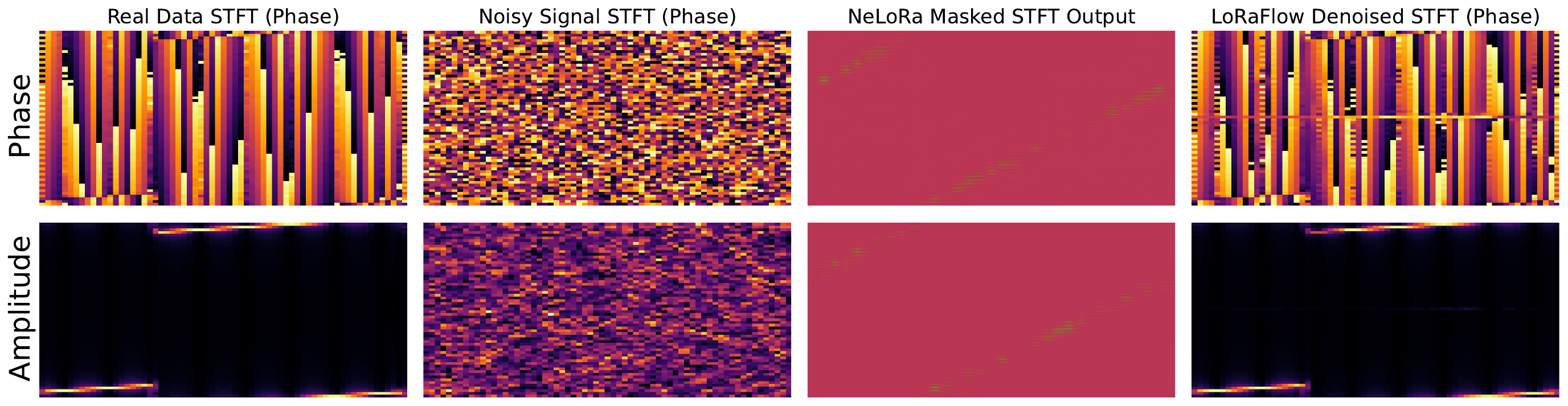}
    \vspace{-0.1in}
    \caption{This figure shows a side-by-side comparison for both the amplitude and the phase of a sample from the NELoRa dataset, the sample at-reception (noisy), NELoRa's masked output which is their version of denoising, and our output.}
    \vspace{-0.15in}
 \label{fig:qualitative_nelora_vs_us}
\end{figure*}

\vspace{-0.05in}
\subsection{Qualitative Comparison of Signal Reconstruction}
Figure \ref{fig:qualitative_nelora_vs_us} presents a comparative analysis of signal reconstruction quality at different processing stages, highlighting LoRaFlow's advanced signal recovery capabilities. The original data displays distinct chirp patterns in both phase and amplitude domains, typical of LoRa modulation. In contrast, the noisy signal shows substantial degradation, with chirp structures becoming indistinguishable, particularly in the amplitude domain. NELoRa's masked output slightly improves upon this noisy signal, yet it does not completely restore the original signal structure, especially in the phase domain. On the other hand, LoRaFlow's output closely mirrors the original signal in both domains, achieving a highly accurate reconstruction of chirp patterns. Impressively, LoRaFlow reconstructs the phase with near perfection, despite phase generally being challenging to accurately restore.

The superior reconstruction quality of LoRaFlow, particularly in preserving phase information, stands out. This ability to maintain phase accuracy is crucial for correct signal decoding in radio communications. LoRaFlow's capability to reconstruct detailed signal structure from heavily noisy environments highlights its potential to considerably enhance the operational range and reliability of LoRa networks in challenging conditions.

\begin{figure}[htbp]
    \vspace{-0.05in}
    \centering
    \includegraphics[width=0.8\columnwidth]{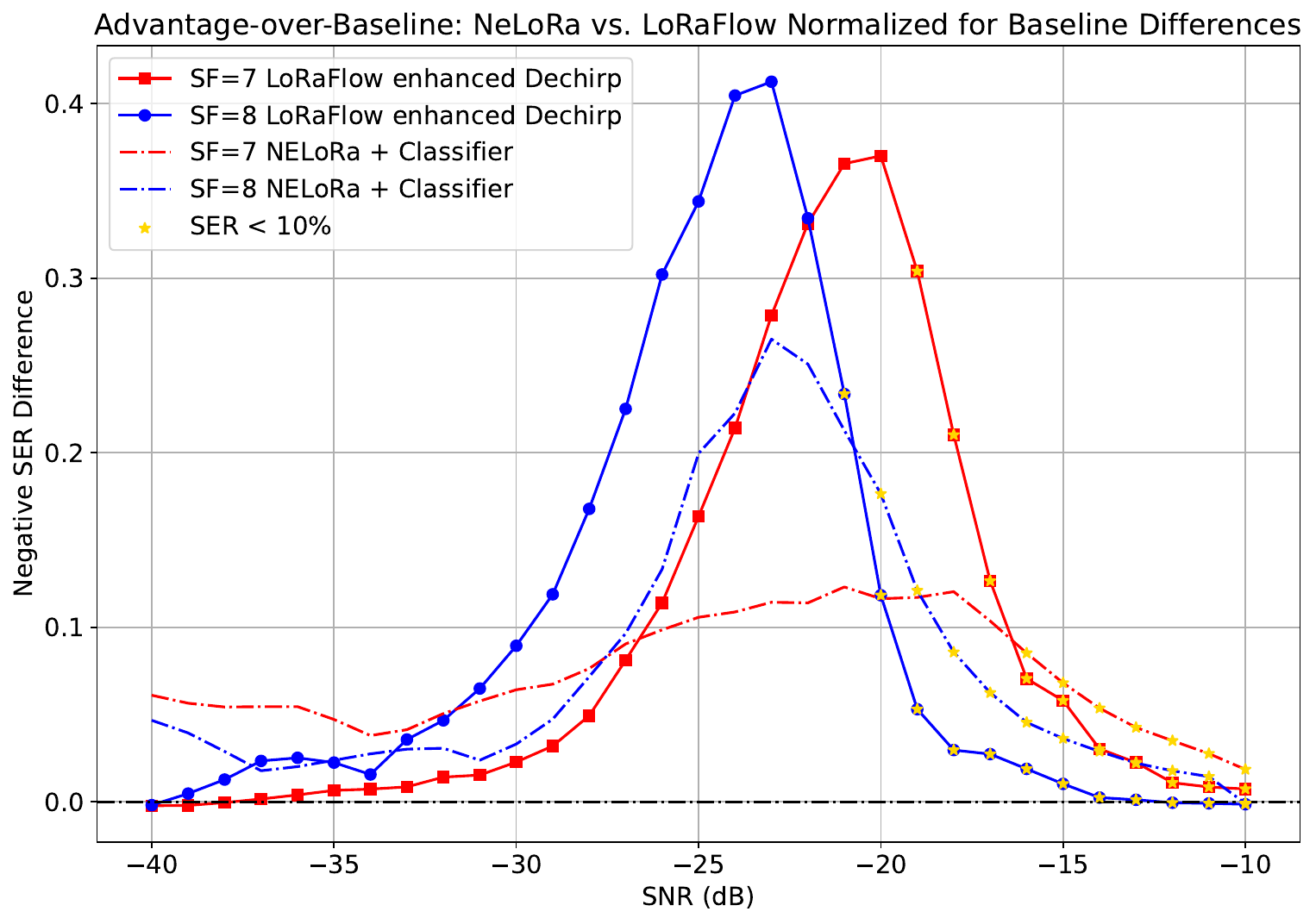}
    \vspace{-0.1in}
    \caption{This plot shows the negative signal error rate advantage over each method's respective baseline. Values represent an error rate \textit{decrease} over the baseline (higher is better). Plot is corrected for differences in baselines.}
    \label{fig:us_vs_lora_ser_advantage}
    \vspace{-0.2in}
\end{figure}

\subsection{Quantitative Comparison with NELoRa}
Figure \ref{fig:us_vs_lora_ser_advantage} provides a quantitative comparison between LoRaFlow and the state-of-the-art NELoRa method, utilizing the metric of Symbol Error Rate (SER) advantage over their respective baselines. This analysis demonstrates that LoRaFlow consistently outperforms NELoRa across various SNRs for both SFs 7 and 8. The performance advantage is notably significant in the -30 to -20 dB SNR range, where LoRaFlow achieves up to a 0.4 SER advantage over the baseline for SF=8. Both methods exhibit diminished advantages at extremely low (below -35 dB) and high (above -15 dB) SNRs, highlighting the challenges of signal recovery in extreme noise environments and the adequacy of traditional methods in the weaker SNR regions.

Our results are particularly significant as they are based on evaluations using almost the entire dataset, whereas NELoRa's reported results cover only 20\% of the dataset, lending greater statistical significance to our findings which is remarkable as our scenario is significantly more challenging. Additionally, while both models have matched parameter counts, implying similar capacities, NELoRa utilizes a separately trained network for each SF, whereas our approach employs a single model for all configurations.

We attempted to reproduce NELoRa's results using the latest code uploaded to their Github\cite{NeLoRaDataset2023}, however, as of writing all the models published on their Github failed to outperform the baseline. Therefore we had to resort to copying NELoRa's results from the figures in their paper\cite{nelora2021}

To provide a holistic performance summary, we compute the Area Under Curve (AUC) for both methods across all SNRs. Table \ref{tab:auc_comparison} presents these results:

\begin{table}[h]
\centering
\caption{AUC Comparison between LoRaFlow and NELoRa}
\label{tab:auc_comparison}
\vspace{-0.05in}
\begin{tabular}{cccc}
\hline
SF & LoRaFlow AUC & NELoRa AUC & Improvement over NELoRa \\
\hline
7 & \textbf{2.922} & 2.227 & 31.2\% \\
8 & \textbf{3.143} & 2.409 & 30.5\% \\
\hline
\end{tabular}
\vspace{-0.15in}
\end{table}

These AUC values (where higher is better) demonstrate LoRaFlow's significant overall performance improvement over NELoRa, with enhancements of 31.2\% and 30.5\% for SF=7 and SF=8, respectively.

\subsection{Discussion}
While LoRaFlow demonstrates significant improvements over NELoRa for SF 7 and 8, there remain opportunities for enhancement, particularly for higher spreading factors. Our analysis reveals several key challenges:

\begin{itemize}[leftmargin=*,noitemsep,topsep=0pt]
    \item \textbf{Dataset Characteristics:} The NELoRa dataset exhibits substantial class imbalance, especially pronounced in higher SFs. This imbalance potentially favors methods that learn the training set's class distribution, an effect that becomes more significant as the number of classes increases (e.g., 512 for SF 9, 1024 for SF 10). In contrast, LoRaFlow's training on a uniform class distribution, while ensuring unbiased performance, does not at all leverage the dataset's statistics. %
    
    \item \textbf{Computational Constraints:} Memory limitations necessitate progressively smaller batch sizes as SF increases, potentially impacting model optimization for higher SFs. This challenge highlights the need for more efficient training strategies or software optimization solutions to maintain samples seen across all SFs.
    
    \item \textbf{Decoding Strategy:} Our approach uses the standard dechirp algorithm without class bias. While this ensures fairness, it may not fully exploit the model's potential. Learning a classifier on top of LoRaFlow could significantly enhance performance, especially for higher SFs.
\end{itemize}

\section{Conclusion}\label{sec:conclusion}
\vspace{-0.02in}
This paper introduces LoRaFlow, a novel approach to LoRa signal reconstruction using rectified flow. Our method demonstrates significant improvements over existing techniques, particularly at low SNRs and for lower spreading factors. LoRaFlow's ability to recover high-fidelity signals from extremely noisy inputs pushes the boundaries of reliable long-range, low-power communication. By maintaining compatibility with existing LoRa infrastructure, our approach offers a practical path to enhancing IoT network performance without overhauling current systems. Future work will focus on addressing challenges at higher spreading factors and exploring the integration of learned classifiers to further boost performance. LoRaFlow represents a significant step forward in robust IoT communications, opening new possibilities for deploying IoT networks in challenging environments.

\section{Acknowledgements}
High Performance Computing resources provided by the High Performance Research Computing (HPRC) core facility at Virginia Commonwealth University (https://hprc.vcu.edu) were used for conducting the research reported in this work. 

\bibliographystyle{IEEEtran}

\bibliography{spaghettibib}

\end{document}